\documentstyle [aps,epsfig,float,twocolumn,prb]{revtex}
\begin{document}
\twocolumn[\hsize\textwidth\columnwidth\hsize\csname @twocolumnfalse\endcsname \draft
\title{Charge dynamics and Kondo effect in single electron traps in field effect transistors}
\author{I. Martin and D. Mozyrsky\\
Theoretical Division, Los Alamos National Laboratory, Los Alamos, NM 87545, USA}
\maketitle
\begin{abstract}
We study magneto-electric properties of single electron traps in metal-oxide-semiconductor field
effect transistors. Using a microscopic description of the system based on the single-site
Anderson-Holstein model, we derive an effective low energy action for the system.  The behavior of
the system is characterized by simultaneous polaron tunneling (corresponding to the charging and
discharging of the trap) and Kondo screening of the trap spin in the singly occupied state.  Hence,
the obtained state of the system is a hybrid between the Kondo regime, typically associated with
single electron occupancy, and the mixed valence regime, associated with large charge fluctuations.
In the presence of a strong magnetic field, we demonstrate that the system is equivalent to a two
level-level system coupled to an Ohmic bath, with a bias controlled by the applied magnetic field.
Due to the Kondo screening, the effect of the magnetic field is significantly suppressed in the
singly occupied state.  We claim that this suppression can be responsible for the experimentally
observed anomalous magnetic field dependence of the average trap occupancy in ${\rm Si-Si0_2}$
field effect transistors.
\end{abstract}
  \vspace{+0.5 cm}]
\section{INTRODUCTION} Experimental techniques probing dynamics  of a
few state quantum system have been of great interest recently as they may provide new prospects for
the development of electronics and computing. Examples of such systems and techniques in solid
state include quantum dots~\cite{weil}, superconducting qubits~\cite{mahlin}, magnetic resonant
force microscopy~\cite{sidles}, and single electron traps in a metal-oxide-semiconductor field
effect transistor (MOSFET)~\cite{ralls,cobden,rts,xiao,ghi}.

This latter system, which is a subject of this work, consists of a defect located near the
oxide-semiconductor interface of a MOSFET (Fig 1(a)). The tunneling between the defect and the
2-dimensional electron gas (2DEG) in MOSFET inversion layer manifests itself as a random telegraph
signal (RTS) in the transport current through the MOSFET. When a certain trap energy level (whose
energy can be controlled by the gate voltage) crosses the chemical potential of the 2DEG, electrons
can hop between the level and the conduction channel, thus charging and discharging the trap. For a
sufficiently small MOSFET's this leads to the sudden switching in the resistivity of the conduction
channel and hence to RTS in the transport current.

The charge dynamics of the trap exhibits a number of features consistent with dynamics  of a
two-level system (TLS) coupled to an Ohmic environment~\cite{cobden}. In particular, the
experimental dependence of tunnel rates on the TLS bias (controlled by the gate voltage) and
external temperature was found to agree with those calculated for the spin-boson
Hamiltonian~\cite{leggett}. However there is a number of RTS properties that cannot be explained
based on the TLS phenomenology.

First of all, the RTS is observed to occur on a relatively large, {\it millisecond-to-second} time
scale, which seems to indicate that the observed traps are positioned sufficiently far, at
distances 20-25 \AA \ from the 2DEG. At the same time, the direct electrostatic measurements of
trap positions indicate that this is usually not the case -- the traps are located only a few \AA \
from the inversion layer~\cite{ghi}.

Another aspect that clearly lies outside the scope of TLS phenomenology has recently been revealed
in the experiments by Xiao {\it et al}~\cite{xiao}, where the dependence of the RTS on applied
magnetic field was studied.  In these experiments, from the gate voltage and the magnetic field
dependence of RTS, it was determined that the most likely origin of RTS was the random switching
between (spin-1/2, neutral) and (spin-0, negatively-charged) states of the trap. In particular, at
sufficiently high temperatures (above several degrees K), the probability of the filled state of
the trap was rapidly decreasing with applied magnetic field, consistent with the simple model of a
spin-1/2 empty state and a singlet filled state. However, at lower temperatures significant
deviations from the simple paramagnetic behavior was observed, possibly indicating quenching of
trap's magnetic moment~\cite{xiao}. Such magnetic behavior appears to be consistent with the Kondo
effect.
\begin{figure}[htbp]
\begin{center}
\includegraphics[height=3.2 in, width=2.2 in, angle=90]{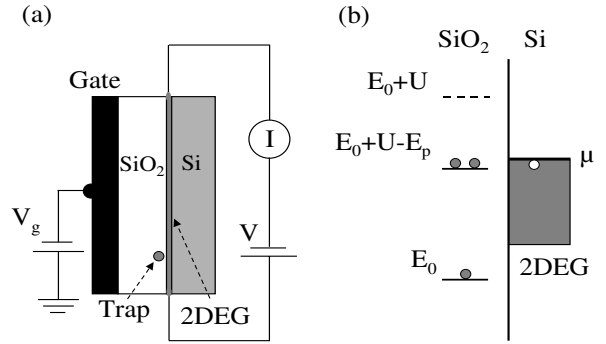}
\vspace{+0.25cm} \caption{(a) Schematics of traps in MOSFETs; (b) A diagram for the trap energy
levels. Coupling to optical phonons shifts the twice occupied level.} \label{fig:setup}
\end{center}
\end{figure}
However, it is well known that the Kondo screening is only effective for sufficiently strong
hybridization of the localized state with the continuum. Given the extremely slow observed tunnel
rate, one should conclude that the Kondo temperature should be negligibly small, thus ruling out
the possibility of the Kondo effect explanation.

The purpose of this work is to show that the slow charge dynamics and the Kondo effect are in fact
{\em not} mutually exclusive, if we take into account the strong electron-optical phonon coupling
in the oxide layer of a Si MOSFET.  In our previous work~\cite{polaron} we have demonstrated that
this coupling is the origin of exponential renormalization in RTS timescales and explains the
inconsistency between expected and observed positions of traps. In the present work, starting from
the microscopic description of the system based on Anderson-Holstein model, we derive an effective
low energy action for the charge {\em and spin} dynamics of the trap. We find that the
low-temperature behavior of the system is characterized by simultaneous polaron tunneling
(corresponding to the charging and discharging of the trap) and Kondo screening of the trap spin in
the singly occupied state. In the presence of a strong magnetic field, we map our system onto a two
level-level system (TLS) strongly coupled to an Ohmic bath, with a bias controlled by the applied
magnetic field. Due to the Kondo screening, the bias introduced by the magnetic field is
significantly suppressed compared to the bare Zeeman energy. We claim that this suppression can be
responsible for the experimentally observed~\cite{xiao} anomalous magnetic field dependence of the
average trap occupancy in ${\rm Si-SiO_2}$ field effect transistors. It is crucial that the Kondo
temperature need not be small as it is controlled by the {\em bare} hybridization matrix element
between the localized and conduction electrons, i.e., it is not affected by the polaronic
slow-down.
\section{MODEL}

To describe the electronic part of the trap-channel system we use the Anderson Hamiltonian:
\begin{eqnarray}
H_{\rm A} = \sum_{k\, \sigma} E_{k} c_{k\,\sigma}^\dag c_{k\,\sigma} +  \sum_\sigma
E_0 d_\sigma^\dag d_\sigma + Un_\uparrow n_\downarrow\nonumber\\
+\sum_{k\, \sigma}T_{cd}(c_{k\,\sigma}^\dag d_\sigma + d_\sigma^\dag c_{k\, \sigma})\,
,~~~~~~~~\label{a1}
\end{eqnarray}
\noindent where $c_{k\,\sigma}^\dag(c_{k\,\sigma})$ and $d_\sigma^\dag(d_\sigma)$ are
creation(annihilation) operators of electrons in the conduction band and at the localized defect
orbitals respectively, $n_\sigma = d_{\sigma}^\dag d_{\sigma}$ ($\sigma =  \uparrow,\,\downarrow$).
$U$ and $T_{cd}$ are the on-site Coulomb energy and the hybridization matrix element respectively.
The coupling strength between the localized states and the 2DEG can be characterized by
$\Gamma=\pi\nu T_{cd}^2$, where $\nu$ is the density of states in the 2DEG.  The single particle
level of the trap is assumed to be positioned deep below the chemical potential of 2DEG, $E_0<\mu$,
as shown in Fig. 1(b). For simplicity we will assume that the conduction band is symmetric and set
$\mu=0$.

Another interaction that we include in our model is due to optical phonons in the Si MOSFET oxide
layer.  ${\rm SiO_2}$ is a polar crystal and known to exhibit strong electron-optical phonon
coupling leading to formation of polaronic states. We incorporate this electron-phonon coupling in
our model as
\begin{eqnarray}
H = H_{\rm A} + \lambda \left(\sum_\sigma n_\sigma-1\right) {\hat x} + {{\hat p}^2\over 2m} +
{m\omega_0^2{\hat x}^2\over 2}\, .\label{a2}
\end{eqnarray}
\noindent Here $x$ and $p$ are displacement and conjugate momentum of local optical phonon of
frequency $\omega_0$ ($\omega_0\sim 50$ meV at ${\rm Si-SiO_2}$ interface), $m$ is the phonon
effective mass (of the order of ${\rm SiO_2}$ crystal unit cell mass), and $\lambda$ is the
coupling constant between the excess charge in the trap and the phonon. We assume the trap state
with one electron is neutral and therefore introduce the off-setting $-1$ term in the
electron-phonon interaction in Eq.~(\ref{a2}). The strength of the electron-phonon coupling has
been estimated in Ref.~\cite{polaron} and can be expressed in terms of the polaronic shift $E_p =
\lambda^2/2m\omega_0^2 \sim 1\,{\rm eV}$. In the following we will assume that $U\geq
E_p\gg\Gamma>\omega_0$.
\section{EFFECTIVE ACTION}

We analyze low temperature partition function of the system ${\cal Z} = {\rm Tr}[\exp{-\beta H}]$,
where $H$ is given by Eq.~(\ref{a2}) and $\beta^{-1}$ is smaller than any energy scale in the
system. It is convenient to decouple the $U$ term in Hamiltonian~(\ref{a1}) by means of
Hubbard-Stratanovich transformation by writing the spin-spin interaction as $U n_\uparrow
n_\downarrow = -(U/2)(n_\uparrow - n_\downarrow )^2 + (U/2)(n_\uparrow +
n_\downarrow)$~\cite{shriffer}. The partition function can be then cast in the functional integral
form as ${\cal Z} =$
\begin{eqnarray}
&&\int {\cal D}X\,{\cal D}Y \exp{-\int_0^\beta d\tau \left[{M{\dot X}^2\over 2} +
{(X-\Delta)^2\over 4
E_p}+ {Y^2\over 2U}\right]}\nonumber\\
&&\times\langle{\cal T} {\rm e}^{-\int_0^\beta d\tau H_\uparrow [X(\tau)+Y(\tau)]}\rangle\,
\langle{\cal T} {\rm e}^{-\int_0^\beta d\tau H_\downarrow [X(\tau)-Y(\tau)]}\rangle\, ,\label{a3}
\end{eqnarray}
\noindent where $H_\sigma[Z(\tau)]= Z(\tau)\,d_\sigma^\dag(\tau) d_\sigma(\tau)$, $\Delta = E_0 +
(U/2)+2E_p$.  Here the explicit ``time'' dependence of $d_\sigma^\dag (d_\sigma)$ is defined by the
interaction representation with respect to kinetic + tunneling terms in the Hamiltonian~(\ref{a1}).
In Eq.~(\ref{a3}) we have introduced $X = \lambda x + E_0 + (U/2)$ as a coordinate of the
oscillator in units of energy and $M=m/\lambda^2$ as the oscillator's mass in the corresponding
units. The scalar field $Y$ is conjugate to the spin of the trap, ${\cal T}$ stands for
time-ordering and the brackets $\langle\cdot\rangle$ denote averaging with respect to electronic
degrees of freedom. The averaging of the ordered exponents in Eq.~(\ref{a3}) can be done by
following the prescription of Refs.~\cite{hamann,yu} based on the theory of singular integral
equations. Here we provide the result: the functional integral in Eq.~(\ref{a3}) can be cast in the
form ${\cal Z} = \int {\cal D} [X, Y]\exp{[-\int_0^\beta d\tau ({\cal V}+{\cal K})]}$, where
functionals ${\cal V}$ and ${\cal K}$ represent potential and kinetic energy of a particle moving
in two-dimensional (X,Y) plane:
\begin{mathletters}
\label{a4}
\begin{eqnarray}
&&{\cal V} = {(X-\Delta)^2\over 4E_p} + {Y^2\over 2U} + V(X+Y) + V(X-Y)\, ,\nonumber\\
&&V(Z) = {Z\over 2}- {Z\over\pi}{\rm tan}^{-1}\left({Z\over\Gamma}\right) + {\Gamma\over
2\pi}{\log}\left[ 1+\left({Z\over\Gamma}\right)^2\right]\, ;\\\nonumber
&&{\cal K} = {M{\dot X}^2\over 2} + K(X+Y) + K(X-Y)\, ,\\
\label{a4a} &&K(Z) ={1\over \pi^2}\int_0^\tau d\tau^\prime {\log}(\tau-\tau^\prime)\,{dZ(\tau^\prime)\over d\tau^\prime} {d\over d\tau}\\
\nonumber &&~~~~~~~~~~~~~~~~~~\times\left[{Z(\tau)\over Z^2(\tau) -
Z^2(\tau^\prime)}\,{\log}{\Gamma^2 + Z^2(\tau)\over \Gamma^2 + Z^2(\tau^\prime)} \right]\,
.\label{a4b}
\end{eqnarray}
\end{mathletters}

The potential energy of the ``particle'' is presented in Fig. 2(a). It has four local minima.  The
two minima on the $X$ axis (at $X_2=\Delta-4E_p$ and $X_0=\Delta$ for $U, E_p \gg \Gamma$)
correspond to different charge occupations of the trap, i.e., by 2 and by 0 electrons.
\vspace{0.2cm}
\begin{figure}[htbp]
\begin{center}
\includegraphics[width=3.3 in]{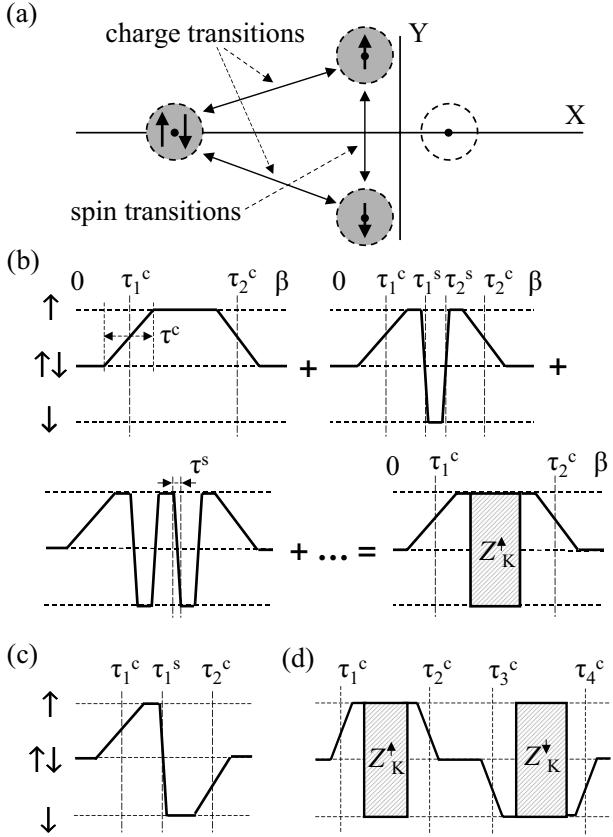}
\vspace{0.65cm} \caption{(a) Local minima for the effective potential in Eq.~(\ref{a4a}); (b) A
classical path contributing to the transition amplitude between a 2 electron and a spin-up electron
states; (c) Example of a path containing a spin-kink "monopole". Such trajectories do not
contribute to the transition amplitudes; (d) Trajectory that involves multiple charge-kinks}
\label{fig:action}
\end{center}
\end{figure}
\vspace{-0.25cm} Up to an additive constant (which we have dropped in Eq.~(\ref{a3})) potential
energies of these minima are ${\cal V}_2 = 2E_0+U-E_p$ and ${\cal V}_0 = -E_p$. The two minima off
the $X$ axis (at $X_\uparrow =\Delta-2E_p$, $Y_\uparrow = U$ and $X_\downarrow =\Delta-2E_p$,
$Y_\downarrow =-U$) are degenerate in the absence of external magnetic field with ${\cal
V}_\uparrow = {\cal V}_\downarrow = E_0$ and obviously correspond to the occupation of the trap by
1 electron with up and down spin, respectively. As can be seen directly from
Hamiltonians~(\ref{a1},\ref{a2}), the minima can be interpreted as zero temperature free energies
of the trap occupied by 2, 0 and by 1 electrons in the limit of vanishingly small $\Gamma$.  In the
RTS experiments the traps are filled with either 1 or 2 electrons~\cite{xiao}, and therefore we
assume that ${\cal V}_\uparrow \simeq {\cal V}_\downarrow \simeq {\cal V}_2 < {\cal V}_0$\ (or
$E_0\simeq 2E_0+U-E_p < -E_p$).
\section{EVALUATION OF FUNCTIONAL}

The kinetic part of the energy in Eq.~(\ref{a4}) will force the ``particle'' to tunnel between the
minima. Because  the energy of the empty state (denoted by the blank circle in Fig. 2(a)) is
greater than that for the other three states (the shaded minima in Fig. 2), it will be effectively
decoupled from those states.  (Transition amplitudes to the empty state will contain exponentially
small prefactors $\exp{-\beta ({\cal V}_0-{\cal V}_2)} \ll 1$.) An example of the ``classical''
trajectory for a transition between the two electron state and spin-up state is presented in Fig.
2(b). Let us first write down explicitly the transition amplitude that corresponds to the first
diagram in Fig. 2(b). In doing so we must, in principle, determine the classical trajectory for the
effective action given by Eq.~(\ref{a4}) between the two local minima corresponding to the singlet
and spin-up states. Solving the resulting classical equation of motion, however, is difficult.
Instead we approximate the classical path by a piece-wise linear kink trajectory, as shown in Fig.
2(b)~\cite{hamann,yu}. Duration of each ``charge'' kink (i.e., transition between the 2 electron
state and a 1 electron state), which we denote by $\tau^c$, can be determined by minimizing the
effective action for a single linear kink trajectory with respect to $\tau^c$. Upon substitution of
such linearized trajectory in Eq.~(\ref{a4}) the transition amplitude of interest can be written as
\begin{eqnarray}
\xi_c^2\int_0^\beta {d\tau_2^c \over \tau^c}\int_0^{\tau_2^c}{d\tau_1^c \over \tau^c}\ {\rm
e}^{\delta_c(\tau_1^c-\tau_2^c) -\alpha_c\log{\tau_2^c-\tau_1^c \over \tau^c}}\ ,\label{a50}
\end{eqnarray}
\noindent where parameters $\xi^c$, $\alpha^c$ and $\tau^c$ are
\begin{mathletters}
\label{a5}
\begin{eqnarray}
&&\alpha_c = \left({2\over\pi}\,{\rm tan}^{-1}{U\over \Gamma}\right)^2\,,\\
\label{a5a} &&\xi_c = \left({\Gamma\over U}\right)^{1\over 2}\,\exp{\left(
-{2E_p\over \tau^c\omega_0^2}\right)}\,,\\
\label{a5b} &&\tau^c \simeq {\sqrt{6}\over\omega_0}\, . \label{a5c}
\end{eqnarray}
\end{mathletters}
\noindent In the above expression for the transition amplitude the kinks at time moments $\tau^c_1$
and $\tau^c_2$ interact logarithmically and $\alpha_c$ is dimensionless coupling strength. The
fugacity $\xi_c$, which is essentially the probability for a kink to occur at a given time, is
suppressed by an exponentially small quantity in Eq.~(\ref{a5a}) as a consequence of the oscillator
having a finite mass. This suppression of charge tunneling is in agreement with the our previous
result~\cite{polaron}, where we argued that the inconsistency between the observed and expected
tunnel rates in the RTS systems is due to the presence of a local phonon strongly coupled to the
trap charge. The quantity $\delta_c$ in the above equation is a bias between one- and two-electron
states, $\delta_c = {\cal V}_2 - {\cal V}_{\uparrow,\downarrow} = E_0 + U - E_p$.

Next let us look at the transition amplitude that involves spin transitions. In Fig. 2(b) this
amplitude can be singled out as an effective self-energy part of the diagrams that involve $2n,\
(n=0, 1, ...)$ kinks and anti-kinks. Applying the same procedure, i.e.,  assuming that all the
trajectories are piece-wise linear, evaluating their contribution to the  effective action in
Eq.~(\ref{a4}) and minimizing the action with respect to the width of the spin kinks $\tau^s$, we
obtain that the spin transition amplitude, ${\cal Z}_K^\sigma$ in Fig. 2(b), is given by
\begin{eqnarray}
{\cal Z}^{\sigma}_K(\tau^c_2-\tau^c_1) = &&\sum_n \xi_s^{2n}\int_{\tau^c_1}^{\tau^c_2}
{d\tau_{2n}^s \over \tau^s}...\int_{\tau^c_1}^{\tau^s_3}{d\tau^s_2 \over
\tau^s}\int_{\tau^c_1}^{\tau^s_2}{d\tau^s_1
\over \tau^s}\nonumber\\
&&\times\exp\left[{-\sigma\delta_s\sum_i{(-1)^i \tau^s_i}+{\sigma\delta_s\beta\over 2} }\right]\nonumber\\
&&\times\exp\left[ \alpha_s \sum_{j>k}^{2n}{(-1)^{j-k}\log{\tau^s_j-\tau^s_k \over
\tau^s}}\right]\label{a6}
\end{eqnarray}
\begin{mathletters}
\label{a7}
\begin{eqnarray}
&&\alpha_s = 2\left({2\over\pi}\,{\rm tan}^{-1}{U\over \Gamma}\right)^2\,,\\
\label{a7a} &&\xi_s = {\Gamma\over U}\, ,\\
\label{a7b} &&\tau^s \simeq {3 \alpha_s\over U} \simeq  {6\over U} \, . \label{a7c}
\end{eqnarray}
\end{mathletters}
Here, $\delta_s = g\mu_B B$ is the energy splitting between the bare spin-down and spin-up states
due to the applied magnetic field $B$~\cite{B}. The transition amplitude ${\cal Z}_K^\sigma$ in
Eq.~(\ref{a6}) is of the same  form as the grand partition function of a Coulomb gas of $\pm$ kinks
positioned along a straight line of length $\langle\tau^c_{2}-\tau^c_{1}\rangle$ interacting
logarithmically with the coupling strength $\alpha_s$ and chemical potential defined as ${\rm
e}^{-\mu}=\xi_s$ ~\cite{anderson}. In order to evaluate the full transition amplitude in Fig. 2(b)
we must, in principle, take into account that the charge kinks at the ends of the interval, i. e.,
at ``positions'' $\tau_1^c$ and $\tau_2^c$, interact (also logarithmically, as can be seen from the
effective action in Eq.~(\ref{a4})) with the spin kinks at positions $\tau_j^s$. One can see,
however, that this interaction is effectively weak. Indeed, kinks at $\tau_j^s$ are coupled
strongly with each other and tend to form close pairs (dipoles) of average effective ``size'' of
the order $d\sim\tau^s/(2-\alpha_s)\sim 1/\Gamma$, while the dipoles are separated, on average, by
distances of the order of $l\sim\tau^s/\xi_s^2\sim U/\Gamma^2\ $~\cite{anderson}. The kinks at
$\tau_i^c$ are separated from those at $\tau_j^s$ by a distance (time) at least of the order
$\tau^c\sim 1/\omega_0$, which physically represents the time needed for the formation of the
dressed electron-phonon state in the trap. Thus, the interaction energy for a kink at $\tau_1^c$
with a dipole at $\tau_1^c+\tau$ is of the order $d/\tau$. Summing over the dipoles (with the
closest dipole located at distance of the order $\tau^c$ from the charge kink at $\tau^c_1$ and the
farthest dipole roughly at $\tau^c_2$) one finds that the charge kink--dipole interaction energy
for the transition amplitude in Fig. 2(b) is $\sim (d/l)\int_{\tau^c}^{\tau^c_2-\tau^c_1}
d\tau/\tau \sim (\Gamma/U)\log{[(\tau^c_2-\tau^c_1)/\tau_c]}$. This interaction is of the same
long-range form as that between the kinks at $\tau_1^c$ and $\tau_1^c$, e.g., Eq.~(\ref{a50}),
however parametrically it is much weaker, $\alpha_c\sim 1\gg(\Gamma/U)$.  Therefore, interaction
between the charge kinks and the spin kinks can be neglected.  Also, we do not need to consider
configurations with an odd number of spin kinks between the charge kinks, e.g. Fig 2(c). That is
because the energy of an unpaired kink diverges logarithmically with
$\langle\tau^c_{2}-\tau^c_{1}\rangle$, and hence its formation is not favorable. Thus the
transition amplitude that contains two charge kinks at $\tau_{1,2}^c$ can be written as
\begin{eqnarray}
\xi_c^2\int_0^\beta {d\tau_2^c \over \tau^c}\int_0^{\tau_2^c}{d\tau_2^c \over \tau^c}\ {\cal
Z}_K(\tau^c_2-\tau^c_1){\rm e}^{\delta_c(\tau_1^c-\tau_2^c) -\alpha_c\log{\tau_2^c-\tau_1^c \over
\tau^c}}\ ,\nonumber
\end{eqnarray}
\noindent where ${\cal Z}_K = {\cal Z}_K^\uparrow + {\cal Z}_K^\downarrow$.

We are now ready to extend this expression to a higher order amplitudes. One such amplitude with
four charge kinks at $\tau^c_j$ is shown in Fig. 2(d).  We notice that only charge kinks connecting
the singlet state with the {\it same} spin state interact.  Hence, we assign superscript $\sigma$
to distinguish between different types of charge kinks, $\tau^c_j \rightarrow \tau^{\sigma_j}_j$.
Using the same energetic arguments as before, we can show that the interaction between the spin
kinks in different domains, as well as the interaction between any charge kinks and spin kinks, is
negligible.  Therefore, the transition amplitude can be written as
\begin{eqnarray}
{\cal Z}=&&\sum_{n,\ \sigma_{2j}=\sigma_{2j-1}= \pm 1} \xi_c^{2n}\int_0^\beta
{d\tau_{2n}^{\sigma_{2n}} \over
\tau^c}...\int_0^{\tau^{\sigma_2}_2}{d\tau^{\sigma_1}_1 \over \tau^c}\nonumber\\
&&\times\prod_{j=1}^n {\cal Z}^{\sigma_{2j}}_K(\tau^{\sigma_{2j}}_{2j}-\tau^{\sigma_{2j}}_{2j-1})\nonumber\\
&&\times\exp\left[-\delta_c\sum_{j=1}^{2n} (-1)^j \tau^{\sigma_{j}}_j\right]\nonumber\\
&&\times\exp\left[\alpha_c
\sum_{j>k}^{2n}(-1)^{j-k}\delta_{\sigma_j,\,\sigma_k}\log{\tau^{\sigma_{j}}_j-\tau^{\sigma_{k}}_k
\over \tau^c}\right]\, .\label{a8}
\end{eqnarray}

In zero magnetic field ($\delta_s= 0$), the spin partition function
$Z_K^{{\uparrow}}(\tau-\tau^\prime) = Z_K^{{\downarrow}}(\tau-\tau^\prime)\sim
\exp{[-(\tau-\tau^\prime)F_0]}$, where $F_0$ is the free energy per unit length for the Coulomb gas
described by the classical partition function of Eq.~(\ref{a6}).  In this regime, the full
partition function of Eq.~(\ref{a8}) corresponds to a gas of charges of {\em two flavors}, with
charges at points $\tau_{2j}$ and $\tau_{2j-1}$ having the same flavor and opposite charge, and
only charges of the same flavor interacting with each other. In the presence of applied magnetic
field, the symmetry between the two flavors is broken.
\section{MAGNETIC FIELD DEPENDENCE}

To evaluate the effect of magnetic field on partition function Eq.~(\ref{a8}), we first determine
the dependence of $Z_K^{\sigma}$ on magnetic field.  For that we use the scaling procedure for the
Coulomb gas due to Anderson {\it et al}~\cite{anderson}. The renormalization group equations read:
\begin{mathletters}
\label{a9}
\begin{eqnarray}
&&{d\alpha_s\over d\log{\tau^s}} = -4 \alpha_s\xi_s^2 + O(\xi_s^3) \ ,\\
\label{a9a} &&{d\xi_s\over d\log{\tau^s}}={1\over 2}\xi_s(2-\alpha_s)+O(\xi_s^4) \ ,\\
\label{a9b} &&{d(\delta_s\tau^s)\over d\log{\tau^s}}= (1-2\xi_s^2)\delta_s\tau^s+O(\xi_s^4)\ , \\
\label{a9c} &&{\cal Z}_K[(\beta/\tau^s),\xi_s,\alpha_s] = \exp{(\beta\Delta F)}{\cal
Z}_K[(\beta/{\tilde \tau}^s),{\tilde \xi}_s,{\tilde \alpha}_s]\ .\label{a9d}
\end{eqnarray}
\end{mathletters}\noindent
where $\Delta F = \int_{\tau^s}^{{\tilde \tau^s}}d\log{{\tau^s}^\prime} \xi_s^2/{\tau^s}^\prime$.
For sufficiently strong magnetic field, ($\delta_s > T_K  = {\tau^s}^{-1}\exp{[-1/\xi_s]}$), the
scaling should be terminated when ${\tilde \delta}_s {\tilde \tau}^s\sim 1$. At this point the
magnetic field becomes very strong (on the renormalized energy scale), while the long-range
coupling constant $\alpha_s$ decreases and so the system moves away from the quantum phase
transition point at $\alpha_s = 2$ into the disordered phase~\cite{anderson}. For mathematical
simplicity we will assume that $2-\alpha_s= 4\xi_s$ in Eqs.~(\ref{a7}).  In such symmetric case the
scaling Eqs.~(\ref{a9a}) and (\ref{a9b}) coincide (up to quadratic terms in $\xi_s$ and
$2-\alpha_s$) and can be easily solved analytically. For $\xi_s\ll 1$ the resulting renormalized
${\tilde \delta_s}$ and ${\tilde\xi_s}$ are
\begin{mathletters}
\label{a10}
\begin{eqnarray}
&& {\tilde \delta_s} = \delta_s \exp[-(1/2)\log^{-1}(\delta_s/T_K) + O(\xi_s)]\ ,\\
\label{a10a} &&{\tilde\xi_s} = \log^{-1}{(\delta_s/T_K)} + O[\log^{-3}{(\delta_s/T_K)}]\
.\label{a10b}
\end{eqnarray}
\end{mathletters}
\noindent It is easy to see that in the rescaled system the long range coupling becomes irrelevant.
A simple perturbative estimate for the average size of a dipole is ${\tilde \tau}^s/({\tilde
\delta}_s{\tilde \tau}^s)^2\sim {\tilde \tau}^s$ and therefore both intra- and inter-dipole
interactions are negligible. ${\cal Z}_K$ can then be readily summed by means of Laplace
transform~\cite{polaron} with a simple result
\begin{eqnarray}
{\cal Z}^{\uparrow}_K(\tau) &&= \exp{\{\tau[F_0^\prime + {\tilde \delta_s} +
O(\log^{-3}{(\delta_s/T_K)})]\}}
\ ,\nonumber\\
{\cal Z}^{\downarrow}_K(\tau) &&\approx {\xi^2_s}\exp{\{\tau[F_0^\prime + {\tilde \delta_s}]\}}
+\exp{\{\tau[F_0^\prime - {\tilde \delta_s}]\}} \ ,\label{a11}
\end{eqnarray}
\noindent where $F_0^\prime$ is magnetic-field-independent free energy and ${\tilde \delta_s}$ is
given by Eq.~(\ref{a10}).   In the long time limit that we are interested in, $Z_K^{\downarrow}$ is
negligible compared to $Z_K^{\uparrow}$.  Hence, in the partition function Eq.~{\ref{a8}}, only the
kinks that correspond to the transition between the singlet and the spin-up states survive.  The
partition function becomes then equivalent to the one for a two-level system coupled to an Ohmic
bath\cite{leggett}.  Eq.~(\ref{a11}) together with Eq.~(\ref{a8}) indicate that an external
magnetic field introduces an effective bias for the TLS.  This bias, however, is significantly
reduced (see Eq.~(\ref{a10})) as compared to the bare Zeeman energy of a paramagnetic spin due to
the strong coupling between kinks in Eq.~(\ref{a6}). The partition function of the latter model is
equivalent to that of the Kondo model~\cite{anderson} and, therefore, the suppression of the
magnetically-induced bias of the TLS is essentially the Kondo effect. Indeed, the magnetization in
the singly occupied state can be easily computed as $M \approx\partial \tilde{\delta_s}/\partial
(g\mu_B B) = \frac{1}{2}(1-1/[2\log(g\mu_B B/T_K)])$, consistent with the Bethe ansatz result of
Andrei {\em et al.}\cite{andrei}.
\section{DISCUSSION AND CONCLUSIONS}

It is important that the Kondo energy scale $T_K$ {\em is not} determined by the extremely slow
charge tunneling rate (of the order of $\xi_c^2/\tau^c$). Rather, the fugacity $\xi_s$, which
determines $T_K$ in Eq.~(\ref{a11}), is due to virtual transitions between Fermi sea and the
localized orbital. In the language of the effective action in Eq.~(\ref{a4}) these virtual
transitions correspond to tunneling of the $Y$-field, which, unlike the $X$-field, is massless.
Therefore, as can be seen from Eq.~(\ref{a6}), the fugacity $\xi_s$ does not contain the
exponentially small quasiclassical suppression factor corresponding to the ``penetration under the
barrier'', which strongly reduces the charge fugacity $\xi_c$ (Eq.(\ref{a5})).   As a result the
Kondo temperature ($\sim\exp{[-1/\xi_s]}$) can be quite large leading to significant
renormalization of the magnetic-field-induced TLS bias.

Based on the above presented arguments, we believe that the anomalously weak magnetic field
dependence of the interface trap occupancy observed in the low temperature experiments of Xiao {\em
et al.} can be explained by the Kondo screening of the local moment in the singly occupied state of
the trap, assuming that the Kondo temperature is $T_K\sim 1$ K.  Then, for low temperatures, $T <
T_K$, we expect that the ratio of the probabilities corresponding to single (spin-up) and double
occupancies of the trap should scale as $P_1/P_2 \propto \exp(g\mu_B {\tilde B}/2T)$, where the
effective magnetic field $\tilde B$ is related to the applied magnetic field $B$ as ${\tilde B} = B
\exp(-1/[2\log(g\mu_B B/T_K)])$ for $g\mu_B B > T_K$.

Clearly, further detailed experimental study is necessary to test out theoretical picture for the
Kondo effect in the electrically active defects (traps) in Si field effect transistors.  It is
worth noting, however, that similar physics, that is simultaneous slow charge dynamics and the
Kondo effect, is expected to occur in any other system that has defects located at the interface
between a strongly polar insulator and a conductor.  One possible way to detect the Kondo effect is
to look for low temperature anomalies in the FET channel resistivity.  Another approach is to look
for spectral features (Kondo peak) by direct tunneling through a defect at the SiO$_2$-Si
interface.

In summary, we have derived an effective action for the charge dynamics in a single electron trap
in a Si MOSFET and have shown that in the limit of a strong magnetic field ($g\mu_B B > T_K$) it is
equivalent to a two level system strongly coupled to an Ohmic environment. The effective bias of
the TLS can be controlled by an external magnetic field.  However, the magnetic field dependence of
the trap spin at low temperatures is suppressed due to the Kondo effect. \section{ACKNOWLEDGEMENTS}

We acknowledge useful discussions with E. Abrahams, M. B. Hastings, H. W. Jiang, and M. Pustilnik.
This work was supported by the US DOE. D.M. was supported, in part, by the US NSF, grant
DMR-0121146.


\begin{references}
\bibitem{B} Though external magnetic field was not introduced explicitly in the above discussion, one can
easily see that its only effect on the effective action in Eq.(\ref{a4}) is to shift the spin
up/down minima of the potential energy by $\pm g\mu_B B/2$.
\end{references}
\end{document}